II

# Comprendre l'atmosphère du Soleil et celles des étoiles variables à la fin du XIXe siècle
*La théorie astro-chimique d'Albert Brester (1843-1919) aux débuts de l'astrophysique*

Guy Boistel

## INTRODUCTION

Dans le domaine des sciences physiques, le XIXe siècle voit l'émergence de nouveaux savoirs tels que la thermodynamique, la cinétique et la statistique des gaz, ainsi que de nouvelles techniques comme la spectroscopie notamment, avec les succès connus de Fraunhofer, Bunsen et Kirchhoff. Ces auteurs et leurs travaux posent entre autres questions, celle de la constitution du Soleil, de son évolution et de l'origine de l'énergie solaire[1].

Les théories solaires foisonnent au XIXe siècle, ce que ne montre pas vraiment ou pas du tout l'historiographie récente sur l'histoire de la physique solaire, notamment l'ouvrage de Karl Hufbauer, devenu désormais un « classique », *Exploring the sun*. Dans cet ouvrage, l'auteur tend à faire l'éloge de la science des vainqueurs[2]. Une telle approche conduit à masquer, voire à ignorer une grande partie des débats qui agitent la communauté savante et l'activité scientifique réelle, c'est-à-dire quotidienne, dans un mouvement de construction et de structuration d'une nouvelle discipline émergeant de l'astronomie physique, l'*astro-physique*[3]. C'est souvent en s'intéressant aux seconds couteaux ou aux savants réputés de second ordre, voire aux théories fausses et/ou rejetées, qu'il est possible de se faire une idée plus précise de la construction, du développement et de la diffusion des savoirs en matière d'histoire des sciences[4].

---

1. Stéphane Le Gars, *L'émergence de l'astronomie physique en France (1860-1914) : acteurs et pratiques*, thèse de doctorat en histoire des sciences et des techniques, Centre François Viète, Université de Nantes, 2007. Francis Beaubois, « Comment construire une théorie du Soleil : problèmes épistémologiques et méthodologiques chez Hervé Faye », in G. Boistel, C. Le Lay et S. Le Gars (dir.), *Hervé Faye (1814-1902) ou l'art de la rupture*, Bulletin de la Sabix, n° 55, septembre 2014.

2. Karl Hufbauer, *Exploring the Sun. Solar science since Galileo*, Baltimore, John Hopkins University Press. Voir aussi A.J. Meadows, *Early solar physics*, Oxford, Pergamon Press, 1970, 1993 (2nde éd.); Jean-Louis et Monique Tassoul, *A concise history of Solar and Stellar Physics*, Princeton and Oxford, Princeton University Press, 2004 (chap. 2 à 4, pages sur les variables et les théories de pulsation stellaire ; Brester n'y est nulle part mentionné).

Il est possible de distinguer trois approches différentes de la physique solaire, c'est-à-dire trois tentatives de décrire ce qu'est la photosphère et de comprendre la constitution et la nature des taches solaires ou des éruptions solaires, dans la seconde moitié du XIX$^e$ siècle. Si, à partir des années 1860, tous les scientifiques considèrent à peu près le Soleil comme étant une boule de gaz[5], cette idée doit encore faire son chemin et résister aux diverses théories qui vont se développer.

La première approche est météorologique ou cyclonique : elle privilégie une analogie avec les phénomènes météorologiques terrestres, regardant l'atmosphère solaire brassée dans son entier par des cyclones et de violentes tempêtes, agitée de vortex et de tourbillons, mélangeant la matière solaire. Alliée aux interprétations spectroscopiques, cette approche constitue le courant dominant, un presque paradigme solaire, jusqu'au début du XX$^e$ siècle. Elle est partagée par les grands noms de la physique solaire de la fin du XIX$^e$ siècle : le père Angelo Secchi s.j. (1818-1878), Hervé Faye (1814-1902), Friedrich Zöllner (1834-1882), William Huggins (1824-1910), Norman Lockyer (1836-1920), Egon Von Oppolzer (1869-1907) et Charles A. Young

---

3. Stéphane Le Gars, *op. cit.* Voir aussi : James E. Keeler, "The importance of astrophysical research and the relation of astro-physics to other physical sciences", *Astrophysical Journal*, vol. 7, n° 4, 1897, p. 271-288.

4. Voir par exemple Hugues Chabot, *Enquête historique sur les savoirs scientifiques rejetés à l'aube du positivisme (1750-1835)*, thèse de doctorat en histoire des sciences et des techniques, Centre François Viète, Université de Nantes, 1999 ; Hugues Chabot, « Une théorie fausse et ses avatars, L'explication cinétique de la gravitation de Lesage à la fin du XIX$^e$ siècle », *Sciences et Techniques en Perspective*, série 2, vol. VII, n° 1, 2003, p. 155-172.

5. Albert Brester, *Essai d'une explication chimique des principaux phénomènes lumineux stellaires*, Delft, J. Waltman Jr., 1888 : « Les étoiles seront regardées dans cette étude comme de véritables soleils, c'est-à-dire, comme des masses sphériques de matières gazeuses agglomérées par l'unique effet de leur propre gravitation », p. 3.

(1834-1908) notamment, tous auteurs d'au moins d'un ouvrage important marquant la littérature astronomique de cette époque.

La seconde approche est chimique. Puisque les spectres révèlent la présence de composés chimiques dans les atmosphères des étoiles et que le Soleil est considéré par presque tous les astronomes comme une boule de gaz à haute température, il est assez normal d'aller puiser dans la chimie et la physique des gaz les lois du comportement des atmosphères stellaires. Cette approche se base sur l'idée de dissociation thermique des éléments chimiques se trouvant dans les étoiles et exploite les réactions chimiques extraordinaires observées sur Terre, notamment mises en lumière par le chimiste Henry Sainte-Claire Deville (1818-1881) en 1864-1867[6], puis exploitées par Faye (en 1866), Secchi (en 1870), Lockyer (en 1885) puis par Albert Brester à partir de 1888. Le vulgarisateur de la science l'abbé Théophile Moreux (1867-1954) sera lui aussi porteur en 1902 d'une théorie hyperthermique du Soleil à rattacher aux idées de Secchi et de Brester[7].

Enfin, la troisième approche fait la part belle à la physique et se base sur la thermodynamique des gaz parfaits. Plus formelle, exploitant les connaissances sur le comportement des gaz et établissant les relations d'équilibres hydrostatiques, et l'influence des équilibres gravitationnel et radiatif mis en jeu dans les atmosphères stellaires, cette approche est défendue par les « vainqueurs » de la physique stellaire[8] (avant que l'on ne comprenne les réactions nucléaires qui se produisent au sein du noyau stellaire) : Jonathan Homer Lane (1819-1880), Karl Schwarzschild (1873-1916), Robert Emden (1862-1940), Nikolay Alekseevich Umov (1846-1915), Harlow Shapley (1885-1972) puis Sir Arthur Eddington (1882-1944), et Meghnad Saha (1893-1956) notamment. Bien que les premières équations de l'équilibre stellaire soient établies dès 1870 par Jonathan Homer Lane[9], ces *vainqueurs* de la physique stellaire ne le sont pas par « K.O. technique », loin s'en faut. Avant qu'une « science normale » des intérieurs stellaires s'établisse, la physique solaire passe par divers chemins et tâtonnements qui participent à une meilleure compréhension des phénomènes stellaires ; c'est ce que je souhaite illustrer dans cette étude.

---

6. Voir note *infra*.

7. L'Abbé Théophile Moreux, *Le problème solaire*, Paris, Berteaux, 1900. Voir les recensions de : Georges Mundler, « Abbé Théophile Moreux. Le problème solaire », *Annales de l'observatoire de Lucien Libert*, vol. III, 1902, p. 48-53. Henri Deslandres, « M. L'abbé Moreux. *Le problème solaire* », *Bulletin astronomique*, série I, vol. XIX, 1902, p. 78-80.

8. Voir le récent : Carl J. Hansen, Steven D. Kawaler et Virginia Trimble, *Stellar Interiors. Physical Principles, Structure and Evolution*, Springer, A&A Library, 2004 (2$^{nde}$ éd.). Pour un aspect plus historique : Jean-Louis et Monique Tassoul, *A concise history of Solar and Stellar Physics*, Princeton University Press, 2004.

## I. LE PARCOURS D'UN ASTRO-CHIMISTE

Intéressons-nous donc à un second couteau de la physique solaire et stellaire, qui n'est cité ou mentionné dans aucun des travaux historiques marquants de ces cinquante dernières années sur l'histoire de la physique solaire : Albert Brester. Né en 1843 à Delft, Brester est d'abord un chimiste et un physicien, sans doute professeur à l'Université de Technologie de Delft (il est présenté comme tel dans la revue royale belge *Ciel et Terre*), et sans doute astronome à Delft. Il est mort en 1919. Nous écrivons « Sans doute astronome », car le parcours qui le conduit à s'intéresser à l'astronomie et particulièrement aux étoiles variables, ne nous est pas encore connu ; Brester est toutefois mentionné par le directeur de l'observatoire Royal de Belgique, Paul Stroobant dans son recensement des observatoires mondiaux en 1907[10]. L'observatoire de Brester, localisé au port fluvial du quartier de Hooikade, semble bien modeste. Brester ne semble pas occuper de position institutionnelle plus importante que celle d'un professeur de chimie physique dans un établissement d'enseignement technologique de la Ville de Delft.

Entre 1888 et 1892, Albert Brester publie trois textes qui constituent le cœur de sa théorie solaire. Il ne la retouchera que pour des détails ou des extensions étrangères à son projet initial avant son décès et son œuvre sera publiée de manière posthume par sa fille aînée en 1924[11]. Dans l'état actuel des recherches, Albert Brester est le premier à publier une tentative de théorie générale stellaire embrassant l'ensemble des phénomènes de l'atmosphère du Soleil et des variations d'éclat des étoiles variables rouges. C'est cet aspect qui rend l'œuvre de Brester intéressante et originale et qui motive la présente étude. Il faut en effet attendre 1935 et la théorie de l'astronome d'Harvard Paul W. Merrill, pour trouver une seconde théorie générale des étoiles variables à longue-période que sont les variables rouges[12].

---

9. Jonathan Homer Lane, "On the theoretical temperature of the Sun; under the hypothesis of a gaseous mass maintaining its volume by its internal heat, and depending on the laws of gases as known to terrestrial experiment", *American Journal of Sciences and Arts*, série 2, vol. 50, 1869, p. 57-74. Voir aussi Corey S. Powell, "J. Homer Lane and the internal structure of the sun", *Journal for the history of astronomy*, 19/3, 1988, p. 183-199.

10. Paul Stroobant *et al.*, *Les observatoires astronomiques et les astronomes*, Bruxelles, Hayez Imprimeur, 1907 ; pour Delft, p. 72.

11. Voir la bibliographie d'Albert Brester en annexe à cette étude.

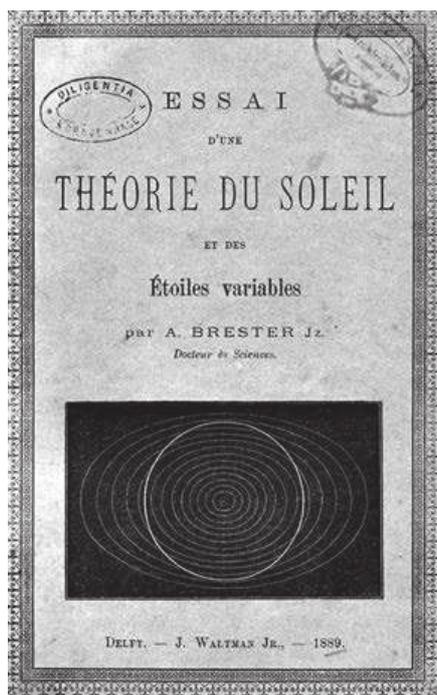

Figure 1. A. Brester,
*Essai d'une théorie du Soleil
et des étoiles variables*, 1889.
(Archives personnelles G. Boistel.)

Brester écrit principalement en français ; le hollandais est réservé aux textes de l'Académie d'Amsterdam au sein de laquelle il semble jouir d'une certaine reconnaissance. Ceci explique sans doute pourquoi les travaux de Brester ne sont cités par aucun des travaux contemporains de Brester sur l'histoire de la physique solaire, ni par Ernst Pringsheim (1859-1917) dans sa *Vorlesungen über die Physik der Sonne*[13]. En revanche, il est abondamment cité et interrogé par deux astronomes français qui trouvent dans ses idées quelque matière à réflexion : Jean Bosler, astronome à l'observatoire de Meudon puis directeur de l'observatoire de Marseille, en 1910 dans son incontournable ouvrage *Les théories modernes du Soleil*[14], en 1923 dans *L'évolution des étoiles*, et encore en 1928 dans le troisième tome de son *Cours d'astronomie* consacré à l'astrophysique[15] ; Jean Mascart, directeur de l'observatoire de Lyon (le seul observatoire français impliqué très tôt dans la recherche sur les étoiles variables), en 1925, écrit une recension intéressante de l'ouvrage posthume d'Albert Brester[16]. La revue de la Société royale belge d'astronomie, *Ciel et Terre*, se fait couramment l'écho de ses travaux et constitue une source de première main. Enfin, Brester a publié quelques articles en langue anglaise, principalement dans *Nature*, auxquels répondent les astronomes anglais Norman Lockyer et Alfred Fowler (1868-1940), deux grands experts en spectroscopie solaire.

---

12. Paul W. Merrill, "Theories of Long-Period variable stars", *Popular astronomy*, 43, 1935, p. 214-221 (7[th] paper). *Id.*, *The nature of variable stars*, N. Y., The Macmillan Company, 1938.

13. Publié à Leipzig, Teubner, 1910.

14. Jean Bosler, *Les théories modernes du Soleil*, Paris, Gaston Doin et Fils, coll. « Encyclopédie Scientifique. Bibliothèque d'astronomie et de mécanique céleste », 1910.

15. Jean Bosler, *Cours d'astronomie. III. Astrophysique*, Paris, Hermann, 1928, p. 204-206.

## II. PERCER LE MYSTÈRE DES ÉTOILES VARIABLES

À l'époque où Brester publie, la plupart des étoiles variables connues sont des étoiles rouges à longue période et à grande amplitude visuelle[17] (hormis les étoiles Algol ou β Persée, δ Céphée par exemple[18]). La description de ces « étoiles changeantes » aux variations de luminosité parfois spectaculaires – elles peuvent disparaître à la vue et réapparaître – a pris une certaine ampleur depuis la fin du XVIII[e] siècle, avec les travaux de James Goodricke et de Nathaniel Pigott[19]. Jérôme Lalande puis François Arago n'hésitent pas à consacrer des pages à ces étoiles dans leurs ouvrages, à parler de leur existence et à les décrire[20]. Mais les étoiles variables (étoiles temporaires ou périodiques ou encore étoiles changeantes) constituent encore des énigmes au milieu du XIX[e] siècle. Seul l'astronome allemand Friedrich Wilhelm Argelander (1799-1875) s'est lancé dans un programme d'observation et de recensement systématiques de ces étoiles, notamment pour des questions de catalogage d'étoiles et d'astrométrie. Argelander en appelle même à l'aide des « amis de l'astronomie[21] » pour développer ce type d'observation et assurer un suivi plus étendu et régulier qu'il ne peut assurer seul.

---

16. Jean Mascart, « Brester. *Le Soleil* », *Rev. Gén. Sci. Pures et Appl.*, t. XXXVI, 1925, p. 84-85.

17. Typiquement une période de l'ordre d'une année et des amplitudes pouvant atteindre et dépasser 3 magnitudes visuelles. L'étoile type Mira de la constellation de la baleine (o Ceti) a pour caractéristiques : période d'environ 331 jours ; amplitude comprise entre les magnitude 2 et 10 ; supergéante rouge de spectre M7IIIe.

18. Algol ou β Persée est actuellement le prototype des algolides, étoiles variables à éclipses. L'étoile δ Céphée est le prototype des céphéides, variables pulsantes d'amplitude notable. Au début du XX[e] siècle, les algolides et les céphéides sont assimilées aux mêmes étoiles ; le concept des pulsations stellaires ne se développe que progressivement ; voir par exemple Michel Luizet, *Les céphéides considérées comme des étoiles doubles, avec une monographie de l'étoile variable δ Céphée*, Annales de l'Université de Lyon, Paris, Gauthier-Villars, 1912.

19. Michael Hoskin, "Goodricke, Pigott and the Quest for Variable Stars", in *Stellar astronomy. Historical studies*, Science History Publications Ltd, 1982, p. 37-55.

20. Voir par exemple : Jérôme Lalande, *Encyclopédie méthodique. Mathématiques*, Panckouke, 1784, t. I : art. *Étoiles*, p. 690-700 ; art. *Changeantes*, p. 340-341. François Arago, *Astronomie Populaire*, Paris, Éditions Barral, 1867, t. I, Livre IX « Des étoiles simples », chapitres XV « Étoiles changeantes ou périodiques » à XXIV « Importance de l'observation des étoiles changeantes » en particulier, p. 386-410.

L'explication des causes des variations des étoiles rouges pose de gros problèmes théoriques. Dans les années 1860, toutes les propositions sont également accueillies. On pense à des croûtes à la surface des étoiles qui occultent en partie la lumière :

> Ces étoiles sont des Soleils, assez refroidis déjà pour que des croûtes d'une grande étendue puissent se développer à leur surface. Le mouvement de rotation apporte alors à notre vue tantôt des parties lumineuses et tantôt des parties obscures. Si la partie encroûtée n'est qu'une minime portion de la surface, l'étoile n'éprouve qu'une variation limitée dans l'éclat de sa lumière, sans disparaître complètement. Que si, au contraire, plus de la moitié de la surface est encroûtée et obscure, il pourra y avoir des époques où l'étoile s'évanouira entièrement[22].

D'autres approches plus analogiques existent. En effet, Hervé Faye a, en 1866, dans un texte encore trop méconnu[23], esquissé un rapprochement entre physique solaire et explications des variations d'éclat des étoiles variables ou étoiles périodiques, jetant ainsi les bases d'une évolution stellaire moderne :

> […] les faits nombreux que nous possédons aujourd'hui nous conduisent à examiner si les étoiles variables et les étoiles nouvelles ne seraient pas autre chose que les états successifs d'un même phénomène dont le Ciel nous offrirait à la fois toutes les phases : les étoiles à éclat constant, les étoiles à faibles variations périodiques, les étoiles à périodes irrégulières, celles qui s'éteignent presque dans leur minima, celles qui cessent de varier pendant un temps plus ou moins long […] enfin les étoiles presque éteintes qui se rallument convulsivement […]. Ne dirait-on pas, je le répète, que ce sont là les phases successives et de plus en plus dégradées de la vie d'une seule et même étoile, phases qui, pour cette étude unique, embrasseraient des myriades de siècles, mais que le ciel nous offre simultanément quand on considère à la fois tous les astres qui y brillent ? De même dans une ville, le spectacle simultané de tous les individus nous fait embrasser d'un seul coup d'œil la succession de toutes les phases qu'un individu pris à part doit traverser jusqu'à sa mort[24].

---

21. Friedrich Wilhelm Argelander, „Aufforderung an Freunde der astronomie", *H.C. Schumacher's Jahrbuch für 1844*, 1844, p. 122-254 ; *Id.*, „Ueber veränderliche Sterne", *Astronomie Nachrichten*, vol. 34, n° 806, 1852, p. 221-222. Sur Argelander et son catalogue d'étoiles : Alan H. Batten, „Argelander and the Bonner Durchmusterung", *J. Roy. Astron. Soc. Can.*, vol. 85, n° 1, 1991, p. 43-50.

22. *Bulletin hebdomadaire de l'association scientifique de France*, t. IV, Paris, Gauthier-Villars, 1868 : septembre 1868, « Catalogue des étoiles changeantes », p. 170-172 ; Explications : tome I du *Bulletin*, p. 236 *sq*.

23. Hervé Faye, « Les étoiles nouvelles et les étoiles variables », *Revue des cours scientifiques de la France et de l'étranger*, n° 38 (18 août), 1866, p. 617-620. Voir G. Boistel, S. Le Gars et C. Le Lay (éd.), *Hervé Faye (1814-1902) ou l'art de la rupture*, éd. cit.

Faye émet l'hypothèse que lors du refroidissement d'une couche gazeuse du Soleil, la température des couches superficielles est suffisante pour que des réactions gazeuses puissent se produire et donner des combinaisons de gaz et de vapeurs nouvelles, peu lumineuses, ou des nuages de particules liquides ou solides à l'incandescence très vive. Pour lui, ces nuages retombent après avoir rayonné dans les couches profondes ; la décomposition qui absorbe une grande quantité de chaleur propage le refroidissement dans les couches profondes qui rompt l'équilibre puis provoque le dégagement de nouveaux gaz. Le processus se répète alors[25].

Curieusement, Hervé Faye ne poursuit pas cette première idée, lui préférant finalement une explication météorologique de l'atmosphère solaire, agitée par de larges cyclones ou vortex, brassant la matière solaire dans son ensemble[26]. C'est d'ailleurs cette approche qui domine chez les principaux astronomes et auteurs de théories solaires dans les années 1870-1900 (Lockyer, Von Oppolzer, Zöllner, Young notamment), à l'époque où Albert Brester entreprend ses recherches sur l'atmosphère solaire.

Dans son œuvre phare de 1892, *Théorie du Soleil*, Brester écrit s'être d'abord intéressé aux étoiles variables (*Essais* de 1888 et 1889)[27]

---

24. Hervé Faye, art. cit., p. 618. En effet, on sait aujourd'hui que lors de sa vie, en fonction de sa masse, une étoile passe par plusieurs états d'instabilité et présente des variations d'éclat caractéristiques de ces bandes d'instabilité que l'on peut dessiner dans le diagramme Hertzsprung-Russell. Voir par exemple John R. Percy, *Understanding Variable Stars*, Cambridge University Press, 2007 ; pour une approche plus technique : Vicki E. Sherwood et Lukas Plaut, *Variable Stars and Stellar Evolution*, Symposium n° 67, I.A.U., Moscow 1974, Dordrecht, D. Reidel Publishing Company, 1975.

25. Hervé Faye, art. cit., p. 619.

26. Voir les études de Stéphane Le Gars et Francis Beaubois dans ce volume.

27. Albert Brester, *Essai d'une explication chimique des principaux phénomènes lumineux stellaires*, Delft, J. Waltman Jr., 1888, 27 p. : « § 1. Explication du phénomène des étoiles avant de penser appliquer sa théorie astro-chimique au Soleil considéré comme une vaste machine thermique. Il s'appuie sur le fait que la spectroscopie montre des signes évidents de combinaisons chimiques dans les spectres des étoiles rouges, que ne montrent pas les spectres des autres étoiles :

> Tandis que notre Soleil et tant d'autres étoiles à spectres analogues sont trop échauffés pour pouvoir contenir des combinaisons chimiques en quantités notables, il n'en est pas de même des étoiles périodiques, qui appartenant au type des étoiles rouges, contiennent d'après les indications de leurs spectres, du moins dans leurs couches extérieures les plus refroidies, des quantités très évidentes de combinaisons chimiques, telles que par exemple des carbures d'hydrogène[28].

Brester reconnaît volontiers ne pas jouer dans la même cour que Secchi, Lockyer ou Young, n'ayant contribué ni aux observations ni aux travaux solaires importants menés en cette fin du XIX[e] siècle. S'il va à l'encontre de l'idée dominante d'une agitation permanente de l'atmosphère solaire, en défendant l'hypothèse d'un Soleil calme troublé par quelques bouffées ou éruptions de chaleur, c'est qu'il espère aussi découvrir « cet aspect saisissant qui, d'après Secchi, est si difficile à trouver, pour convaincre les autres[29] ».

Brester se repose sur une excellente connaissance des écrits de ses contemporains ; il fait une excellente bibliographie et cite très précisément ses sources. Brester apparaît très au courant des travaux et découvertes en matière de spectroscopie et en matière de cinétique des gaz. Il est aussi très au fait des dernières recherches sur le Soleil.

Entrons dans le détail de la théorie de Brester.

---

périodiques et des étoiles temporaires », p. 5-11 ; « § 2. Explication des protubérances solaires », p. 12-14 ». La partie consacrée aux étoiles variables est la plus importante. *Id*., *Essai d'une théorie du Soleil et des étoiles variables*, Delft, J. Waltman Jr., 1889, 48 p. : « I. Les étoiles variables », p. 6-19 ; « II. Le Soleil », p. 20-44.

28. Albert Brester, *Essai d'une explication chimique […]*, éd. cit., p. 5. Les spectres de ces étoiles montrent des bandes moléculaires très fortes pour l'oxyde de titane TiO notamment, ainsi que le groupe CH. Ces spectres correspondent aux spectres de type III chez Angelo Secchi et dans la littérature du XIX[e] siècle où l'on se réfère directement aux types spectraux de Secchi. Voir J.-C. Hénoux, « Molécules dans les atmosphères stellaires », *Journal de physique, Colloque C5a*, supplément au n° 10, t. XXXII, octobre 1971, p. C5a-121-127.

29. Albert Brester, *Théorie du Soleil*, Amsterdam, J. Müller, 1892, p. 7.

## III. BRESTER ET LA PHYSIQUE SOLAIRE

Albert Brester emprunte beaucoup au père Angelo Secchi, lequel propose dans son ouvrage *Le Soleil* [30], indépendamment semble-t-il d'Hervé Faye, une approche chimique des phénomènes régissant l'atmosphère solaire :

> Dans l'intérieur du globe solaire, l'effet dû à la gravitation étant extrêmement considérable, il doit en résulter un état gazeux bien différent de tout ce que nous connaissons sur Terre. D'un côté une pression énorme doit favoriser *l'affinité* ; mais de l'autre la température est tellement élevée, qu'aucune combinaison [chimique] proprement dite ne peut subsister, si ce n'est à la surface où la radiation peut abaisser la température d'une manière suffisante. Les différents corps simples peuvent, en effet, rester l'un en présence de l'autre sans se combiner, malgré leur affinité réciproque : on dit alors qu'ils sont *dissociés*[31].

Secchi s'appuie sur des expériences du chimiste français Henri Sainte-Claire Deville (1818-1881) menées en 1864 et publiées en 1866 dans ses *Leçons sur la dissociation*[32]. Sainte-Claire Deville a en effet montré qu'oxygène et hydrogène refroidis en présence de dioxyde de carbone donnent un mélange tonnant, alors que sans dioxyde de carbone, oxygène et hydrogène se seraient combinés en eau[33]. Secchi reprend l'idée de la dissociation par analogie avec les observations chimiques « *extraordinaires* » faites sur Terre :

> [...] Il y a la plus grande analogie entre les faits que nous venons de rappeler et ceux qui accompagnent les combinaisons chimiques. À l'état de dissociation, les gaz contiennent une certaine quantité de chaleur latente qui devient sensible au moment où la combinaison s'effectue. Toute la chaleur qui disparaît dans la dissociation reparaît dans la combinaison, sans aucune perte [...][34].

---

30. Angelo Secchi s.j., *Le Soleil*, Paris, Gauthier-Villars, 1870 (1$^{re}$ éd.), p. 289-295 en particulier.
31. *Ibid.*, p. 289-290.
32. Henri Sainte-Claire Deville, *Leçons sur la dissociation professées devant la Société chimique, le 18 mars et le 1er avril 1864*, Paris, Imprimerie de C. Lahure, 1866, 126 p. Voir *infra* pour de plus amples détails sur cette « dissociation ».
33. Une autre analogie donnée par Brester est la suivante : le mélange $O_2 + 2\,H_2$ donne un mélange tonnant. Si l'on ajoute une partie 7 ½ d'air, il n'y a plus d'explosion du mélange ; A. Brester, *Théorie du Soleil*, éd. cit., p. 33-35 et la très longue note (x), p. 34-35.
34. Angelo Secchi s.j., *op. cit.*, p. 291.

Cette idée suppose que la chaleur qui dissocie les combinaisons chimiques, dissocie aussi les corps simples pour ne laisser subsister à la fin que les *principes de la chimie universelle* (le concept d'ion est en cours de construction par Svante Arrhenius[35]). Mais Secchi ne développe pas particulièrement cette idée comme se propose de le faire Brester dans ses essais de la période 1888-1892.

## IV. LE SOLEIL TRANQUILLE

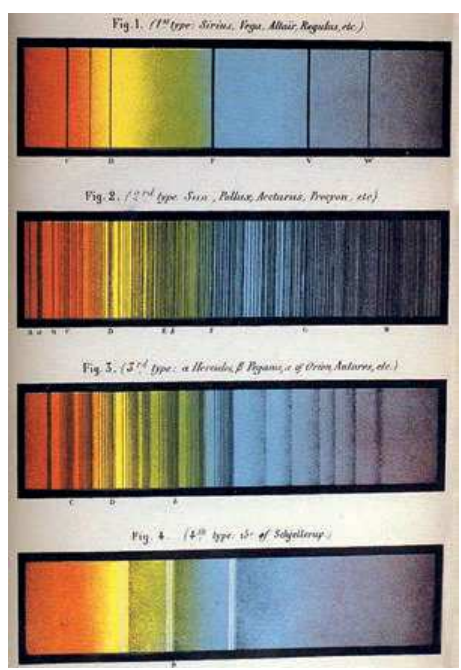

Figure 2. Types spectraux d'Albert Secchi. (Source : A. Secchi, *Les étoiles. Essai d'astronomie sidérale*, Paris, Germer Baillère et C$^{ie}$, 1879, t. I, p. 89.)

Pour Brester, les conditions de température T et de pression P sont compatibles avec un spectre continu jusqu'à la photosphère. Autour de la chromosphère, où la température et la pression sont inférieures, le spectre est naturellement un spectre d'absorption. Les spectres montrent l'existence de molécules (spectres de bandes de type III chez A. Secchi – 7) ; on y voit aussi les éléments chimiques H, He, Ca, Na, Mg, Fe, Ni, Ba et la très fameuse et énigmatique raie observée, celle du pseudo-élément *coronium* (ligne 1474 de Fraunhofer dans la série de raies du fer)[36].

Comme la plupart des principaux auteurs, il partage l'idée que le Soleil est une boule gazeuse, une vaste machine thermique, mais il se démarque en considérant que le Soleil présente une structure stratifiée que met en évidence la spectroscopie, incompatible, écrit-il, avec des mouvements cycloniques ou de brassage de la matière solaire en profondeur comme l'avance notamment Lockyer que Brester critique beaucoup dans sa *Théorie*

---

35. Axel Petit, *Histoire du concept d'ion au XIX$^e$ siècle*, thèse de doctorat en histoire des sciences et des techniques de l'Université de Nantes, Centre François Viète, 2013.

36. George C. Claridge, "Coronium", *Journ. Roy. Astron. Soc. Canada*, 21/8, october 1937, p. 337-346 pour une discussion de près de 70 années autour du coronium.

*du Soleil* de 1892. Ainsi, Brester rejette le recours à toute extension de la météorologie terrestre aux phénomènes solaires et écrit que cette idée est « viciée à sa base ». Il plaide pour un calme de l'atmosphère solaire – « le Soleil tranquille » – et rejette les théories météorologiques solaires de Faye, Secchi, Zöllner et Oppolzer.

> Ma théorie demande en premier lieu que le Soleil soit à l'intérieur relativement tranquille, que du moins, il ne soit pas continuellement bouleversé par toutes ces éruptions, ces explosions et ces tempêtes, qui d'après l'interprétation courante du déplacement des raies spectrales y séviraient sans relâche[37].

Les arguments avancés sont mûris au fil des ans et au gré des découvertes spectroscopiques réalisées à la fin du XIX[e] siècle. En premier lieu, Brester s'approprie les propos de A. Clerke qui, au sujet des spectres des protubérances solaires, se prononce pour « une tempête d'hydrogène dans un calme de calcium[38] ». En effet, si les spectres des protubérances montrent des déplacements de raies spectrales auxquelles on peut appliquer l'effet Doppler, certaines raies ne montrent aucun déplacement, notamment celle du calcium. Brester montre qu'il ne peut y avoir plusieurs gaz en mouvements violents alors que d'autres seraient au repos ; il fait alors une analogie terrestre : « La tranquillité du [pseudo-]coronium serait ici tout aussi impossible que le serait la tranquillité de l'azote dans notre propre atmosphère si l'oxygène s'y mît à souffler[39]. » En second lieu, pour Brester, la stratification des couches solaires développée par Lockyer et Secchi notamment[40] est une preuve convaincante de la (relative) tranquillité du Soleil et il ne comprend pas pourquoi d'autres chercheurs n'adhèrent pas à cette hypothèse. Enfin, le troisième argument en faveur d'un « Soleil tranquille », est l'invariabilité du spectre solaire, hormis quelques événements remarquables rares qui conduisent d'ailleurs, selon lui, à des interprétations souvent divergentes. Brester voit son interprétation relayée par l'astronome jésuite hongrois Julius Fenyi (1845-1927) qui en 1894 s'exprime sur la simplicité de l'explication de Brester :

---

37. Albert Brester, *Essai d'une explication du mécanisme de la périodicité dans le Soleil et les étoiles rouges variables*, Amsterdam, J. Müller, 1908, p. 3.
38. *Ibid.*, note f, p. 4 ; Agnès Clerke, Probl. P. 96 : « Nothing is commoner than the raging of hydrogen storms amid profound calcium calms. »
39. Albert Brester, *Essai d'une explication du mécanisme […]*, éd. cit., p. 4.
40. Références citées dans *ibid.*, p. 5 : N. Lockyer, *The chemistry of the Sun*, p. 304 ; Angelo Secchi, *op. cit.*, t. I, p. 275-293 ; t. II, p. 292, 482 entre autres.

> En considérant ces phénomènes dans leur ensemble et en détail et en cherchant à nous les expliquer nous avons peine à n'y voir qu'un mouvement mécanique : nous opinerions plutôt pour un mouvement apparent, produit par la rapide propagation d'une opération de physique ou de chimie. La théorie de M. Brester, selon qui les protubérances ne sont autre chose qu'une inflammation de ces endroits où les éléments dissociés sont tellement refroidis qu'ils peuvent se réunir, cette théorie, dis-je, nous offre l'explication la plus simple de ces phénomènes […][41].

En résumé, Brester pose comme acquise une stratification stable à l'intérieur du Soleil, et relativement instable à l'extérieure du Soleil en raison du mouvement des masses causé par leur refroidissement. Brester ne peut accepter l'idée des cyclones solaires puisque pour sa théorie, l'atmosphère solaire est garante de la durabilité de la stratification des couches solaires.

## V. LE SOLEIL À L'ÉPREUVE DE L'HYPOTHÈSE DE LA DISSOCIATION CHIMIQUE

S'inscrivant dans la filiation Sainte-Claire Deville-Secchi, Brester pense lui aussi que la dissociation thermique des gaz doit s'opposer à leur refroidissement qui conduirait à une recombinaison chimique qui produirait de la chaleur et procurerait au Soleil une relative stabilité. C'est aussi en creusant cette idée que Brester va chercher à expliquer les variations d'éclat des étoiles, dans des éruptions ou « bouffées » de chaleur sans déplacement de matière autre que la « matière lumineuse ».

> Qu'il me soit permis d'insister beaucoup sur le fait, que dans ce jeu paisible de voiles et de calories se formant et se détruisant tour à tour *nous ne voyons pas des éruptions matérielles de gaz ardents mais seulement des éruptions de chaleur*. Le mouvement des masses y est relativement insignifiant ; et pour peu qu'il existe, il tend à réparer les désordres […] comme étant causés par le refroidissement[42].

---

41. Julius Fenyi, s.j., « Sur deux grandes protubérances de septembre 1893 observées à Kalocsa », *Mem. della Soc. di Spettrosc.*, vol. 23, 1895, p. 28-32, cit. p. 31. Fenyi émet toutefois quelques réserves sur la capacité de cette théorie à s'appliquer aux protubérances, qui ne peuvent se plier à des interprétations de combinaisons chimiques ou physiques en regard des déplacements des raies d'émission dans les spectres.

42. Albert Brester, *Essai d'une explication chimique […]*, éd. cit., p. 9.

Pour Brester, le Soleil est certes une masse gazeuse perpétuellement agitée et bouleversée, dans des proportions moindres que celles des partisans des cyclones solaires, mais une masse gazeuse à haute température et relativement *tranquille*. Cette tranquillité implique une relative invariabilité thermique des diverses parties du Soleil et doit expliquer la formation et la périodicité des taches.

La photosphère toute entière du Soleil subit par le jeu même de la communication de chaleur de l'intérieur vers l'extérieur, et de la perte de chaleur par rayonnement, un mouvement de contraction et de dilatation. Les taches sont alors regardées comme des « trouées » remplies de vapeur absorbante dans la photosphère formées de véritables nuages de condensation plus émissifs. Pour Brester, les phénomènes solaires sont le résultat d'un jeu incessant de condensation et décondensation qui s'apparente à la dissociation chez Faye et Secchi, que Brester nomme un état de *surdissociation*. Cette hypothèse lui paraît ainsi rendre compte « des cas les plus différents de la périodicité comme les effets variés d'une cause unique : la combinaison intermittente à l'extérieur de ce qui était séparé à l'intérieur par la chaleur »[43].

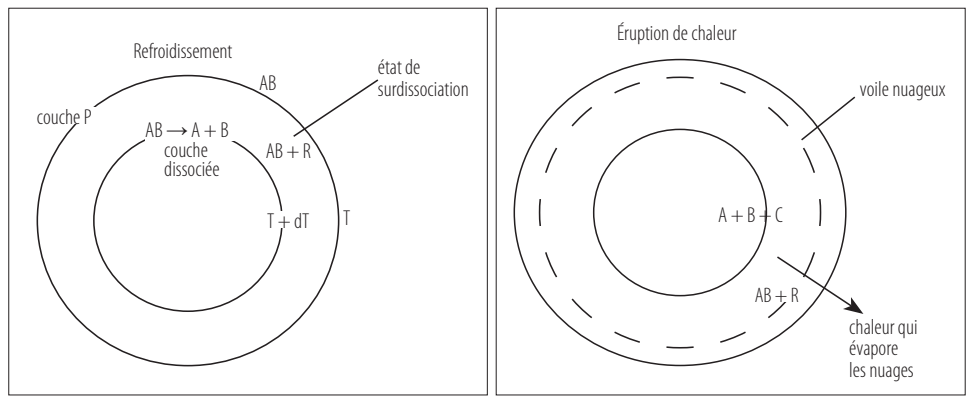

Figure 3a et b. Le mécanisme de *surdissociation*.
(© G. Boistel d'après A. Brester, 1888-1889.)

Examinons le mécanisme astro-chimique imaginé par Brester. Tout refroidissement est rendu impossible et tout refroidissement est inégal. Il fait intervenir : 1°. la condensation des gaz et le dégagement de chaleur latente lors de la recombinaison d'éléments chimiques ; 2°. la combinaison d'éléments chimiques dissociés qui libère leur chaleur de réaction. Brester lève quelques obstacles théoriques en faisant intervenir des molécules étrangères en nombre suffisant qui empêchent des recombinaisons qui s'effectueraient sans elles : c'est l'état de « surdissociation » et Brester en appelle aux expériences de Sainte-Claire Deville. Dans l'atmosphère solaire explique-t-il, si les molécules étrangères sont trop peu nombreuses (ce sont alors les molécules gazeuses libres de la théorie cinétique des gaz), les réactions de recombinaison peuvent survenir brutalement et produire des bouffées de chaleur. L'éruption de chaleur qui se produit dissocie alors les molécules des corps interposés qui se recombinent lors de l'arrêt de la bouffée de chaleur ; et l'on assiste alors à une suite de phénomènes plus ou moins périodiques.

---

43. *Ibid.*, p. 11.

Dans ce mécanisme de surdissociation, lors du refroidissement, la couche P plonge et la distance entre les molécules A et B diminue (figure 3a). On observe alors un mélange A+B+R (molécules étrangères) et de AB. Les molécules A et B s'unissent dans des couches de température plus élevées et contribuent au réchauffement des couches inférieures. La température augmente provoquant la dissociation AB → A+B et propageant les molécules vers l'extérieur (figure 3b) provoquant une « bouffée de chaleur ». Selon la présence massive ou non des molécules étrangères R, l'état de l'atmosphère et des couches solaires peut évoluer :

> Revenons […] à notre couche A+B+R de l'astre, si dans cette couche le pouvoir synthétique encore très petit des molécules A et B ne parvient pas à vaincre l'obstacle des molécules R, cet état surdissocié des molécules A et B pourra durer très longtemps […] *Mais comme ce n'est pas seulement la température encore élevée, mais surtout aussi la présence des molécules R innombrables qui empêche l'union des molécules A et B, une diminution suffisante dans le nombre de ces molécules R aura le même effet qu'une diminution de température […]* Ce seront alors ces molécules R, qui seules auront à accomplir la tâche importante de s'opposer quelques temps par leur condensation au refroidissement de la couche stellaire[44].

La photosphère est donc pour Brester, une mer de nuages où se forment les taches solaires. La bouffée de chaleur vaporise les masses inertes, creuse un trou renforcé par l'absorption causée par la vaporisation ou la production de nuages de gaz absorbant. Brester parvient notamment à expliquer la chaleur plus faible observée pour les taches solaires, pour lesquelles il invoque un plus faible pouvoir émissif.

---

44. Albert Brester, *Théorie du Soleil*, éd. cit., p. 35.

# VI. COMPRENDRE LE COMPORTEMENT PHYSIQUE DE L'ATMOSPHÈRE DES ÉTOILES VARIABLES ROUGES

Brester est le premier auteur à ouvrir un ouvrage sur une théorie du Soleil par un chapitre sur les étoiles variables. Jusqu'alors, ces étoiles sont principalement demeurées dans le domaine observationnel et les essais d'interprétation des variations d'éclats sont restés très modestes. La plupart des astronomes considèrent que les variations d'éclat sont dues soit à des taches qui obscurcissent la surface des étoiles ou bien à des étoiles présentant une face obscure, l'autre étant brillante. En 1784, John Goodricke a bien proposé un mécanisme d'éclipses pour l'étoile β Persée ou Algol[45] mais l'idée d'un corps sombre tournant autour d'un soleil est loin d'être universellement partagée. Brester n'accepte pas la possibilité de considérer l'existence de systèmes binaires à éclipses ou l'hypothèse d'un compagnon obscur tournant à très grande vitesse autour d'une étoile brillante. Ainsi, Brester amalgame le type Algol avec les autres variables. Il ne peut imaginer que l'étoile U Ophiuchi a un compagnon qui tourne en une vingtaine d'heures[46] ! :

> Les autres explications proposées pour le phénomène du type Algol semblent bien peu probables et même peu sérieuses. Elles supposent en effet l'existence d'objets aussi exorbitants que des étoiles toujours obscures d'un même côté et toujours brillantes de l'autre ou bien d'étoiles obscures gravitant autour d'un astre comme U Ophiuchi par exemple dans une période de 20 heures ! […] et il nous faut faire intervenir encore d'autres astres tout aussi obscurs mais perturbateurs pour expliquer les changements que la périodicité présente ;

---

45. Michael Hoskin, "Goodricke, Pigott and the Quest for variable stars", in *Stellar astronomy. Historical studies*, Science History Publications Ltd., 1982, p. 37-55. *Id.*, "Novae and variables before the spectroscope", *Journal for the history of astronomy*, 38/3, n° 132, 2007, p. 365-379.

46. U Oph est même un système triple à éclipses de période P = 1,6773460 jour environ : V.P. Kukarkin *et al.*, *General Catalogue of Variable Stars, 1969 ed.*, t. II, 1970, p. 134-135. Voir aussi : L.P.R. Vaz, J. Andersen et A. Claret, "The eclipsing triple system U Ophiuchi revisited", in W.I. Hartkopf, E.F. Guinan et P. Harmanec (éd.), *Binary Stars as Critical Tools & Tests in Contemporary Astrophysics, Proceedings of IAU Symposium n° 240, held 22-25 August, 2006 in Prague, Czech Republic*, Cambridge University Press, 2007, p. 109-110.

changements qui font retarder par exemple Y Cygni d'une demi-seconde à chaque minimum nouveau[47].

Brester mobilise tous les résultats connus qui vont en ce sens et sonnent comme un coup de grâce donné à l'hypothèse du satellite obscur[48]. Avec la même détermination, il rejette l'hypothèse de l'essaim météoritique avancée par Norman Lockyer pour expliquer les variations d'éclat d'Algol.

Comme Hervé Faye l'avait écrit en 1866, Brester rejette l'appellation « étoiles nouvelles » pour décrire des étoiles dont les variations paraissent soudaines ou irrégulières. Il lui préfère l'appellation d'« étoiles temporaires » dans sa quête d'une théorie qui expliquerait l'ensemble des phénomènes observés :

> Les étoiles nouvelles ne méritent pas ce nom ; leur apparition presque subite n'est qu'une exagération du phénomène ordinaire des étoiles périodiquement variables, lequel répond lui-même à de simples oscillations plus ou moins sensibles dans le phénomène de la production et de l'entretien des photosphères de toutes les étoiles[49].

Dans les années 1880, les variations d'éclat de ces étoiles « changeantes » ou périodiques demeurent donc toujours aussi mystérieuses.

Brester étend son hypothèse de la surdissociation pour expliquer la cause des variations d'éclat des étoiles rouges. Le mécanisme des éruptions de chaleur est celui décrit pour le Soleil. Les nuages produits par le refroidissement de vapeurs jusqu'à leur point de rosée, sont la cause d'une opacité qui amoindrit l'éclat de l'étoile en vertu du sombre voile qui l'enveloppe :

> Dans les étoiles variables, la condensation intermittente des molécules R produit par intervalles à l'extérieur de l'astre des nuages obscurcissants. *Plus ces nuages s'épaississent, plus ils voilent l'éclat intérieur de l'astre, plus aussi ils préparent un maximum nouveau.* Car dès que, l'étoile étant au minimum, la condensation des molécules R en nuages aura atteint une certaine limite, les molécules

---

47. Albert Brester , *Essai d'une théorie du Soleil […]*, éd. cit., p. 16-17. Brester reconnaît qu'il peut se produire des changements périodiques dans les systèmes doubles ; il ne parle pas de mouvement apsidal mais l'idée est là.

48. Albert Brester , *Essai d'une théorie du Soleil […]*, éd. cit., p. 17-19.

49. Hervé Faye, art. cit., p. 620.

A et B, que ces molécules séparaient, cessant d'être suffisamment séparées, se combineront, et produiront de la sorte une éruption de chaleur, qui en évaporant les nuages du minimum restaurera le maximum en rendant mieux visible de nouveau l'intérieur toujours invariablement brillant de l'astre[50].

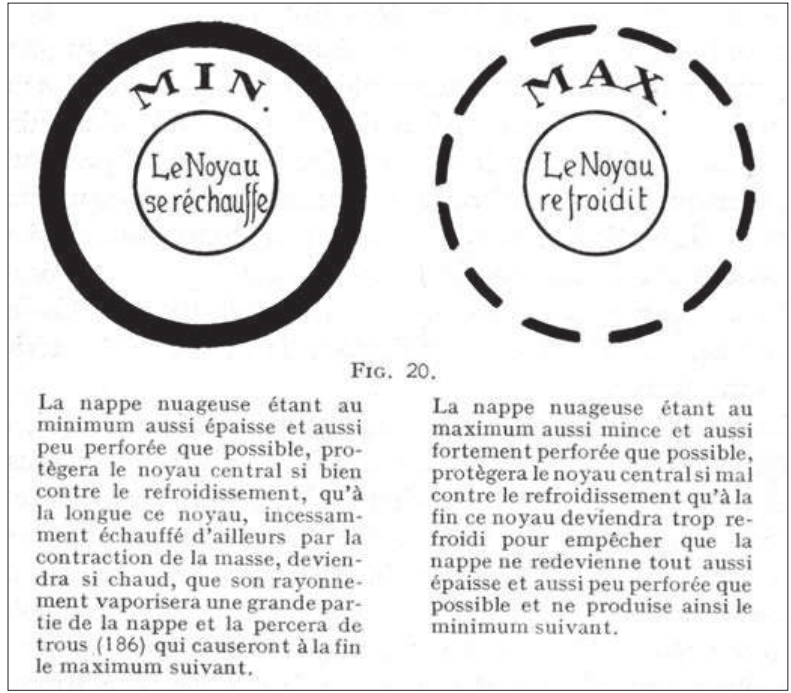

Figure 4. Explication des variations dans les étoiles rouges et le Soleil par A. Brester. (Source : *Le Soleil*, 1924, p. 145. Archives personnelles G. Boistel.)

Pour la plupart des étoiles, remarque Brester, ces changements périodiques passent inaperçus mais pour les étoiles rouges, les combinaisons chimiques mises en évidence par l'analyse spectrale doivent occasionner des changements nettement perceptibles. La stabilité des périodes est le signe d'une composition chimique stable des couches externes. Brester explique que le maximum long et stable observé dans les variations d'éclat d'Algol est dû à un refroidissement des gaz jusqu'au point de rosée. Les maxima et minima secondaires des étoiles du type β Lyrae[51] sont dues à des éruptions de chaleur qui ne conduisent pas la surdissociation jusqu'à son terme mais à un retour à une température différente de la température initiale : « l'énergie chimique des étoiles doit s'opposer d'une manière

---

50. Albert Brester, *Théorie du Soleil*, éd. cit., p. 38.
51. Étoile variable à éclipses mais présentant une courbe arrondie aux maximums de lumière, au contraire des algolides qui présentent souvent un maximum plat ou presque.

intermittente à leur refroidissement […] le phénomène que les étoiles variables nous présente est causé par un mécanisme tout à fait comparable à une horloge dont l'énergie chimique serait le ressort immense et la chaleur produite l'échappement[52]. »

Invoquant ce « jeu paisible de voiles et de calories se formant et se diluant [où] nous ne voyons pas d'éruptions matérielles de gaz mais seulement des éruptions de chaleur », Brester affirme que :

> […] ce qui distingue mon hypothèse chimique, c'est qu'elle est tout à fait générale et qu'en expliquant tous les cas de la périodicité comme les effets variés d'une cause unique : *la combinaison intermittente à l'extérieur de ce qui était séparé à l'intérieur par la chaleur*, elle rend compte en même temps de ce fait capital, écartant toute explication à rotation quelconque, que parmi tous les objets que le ciel nous présente, il n'y en a pas de plus capricieux que les étoiles variables[53].

Cherchant à décrire toutes les étoiles variables sous l'angle de la dissociation thermique des constituants chimiques de l'atmosphère solaire, Brester se ferme ainsi à d'autres tentatives d'interprétations.

## VII. RÉCEPTION ET DISCUSSION DES TRAVAUX D'ALBERT BRESTER

L'étude de la réception des travaux de Brester permet de mieux saisir comment les hypothèses et les idées neuves sur l'atmosphère solaire circulent, et d'observer les pratiques locales de cette nouvelle discipline en construction, l'astrophysique.

La table 1 donne les dix-huit articles de revues, annonçant les travaux de Brester, avec des critiques et ses réponses. On peut y discerner trois époques différentes d'une dizaine d'années chacune environ. La première, de 1888 à 1896, voit les travaux de Brester discutés dans des revues anglaises de haute réputation scientifique, *Nature*, et dans les premières publications spécialisées en astrophysique, l'*Astrophysical Journal* et *Astronomy&Astrophysics* (connus depuis sous les acronymes *ApJ* et *A&A*). De grands noms de l'astronomie prennent le temps de discuter avec Brester : Alfred Fowler, astrophysicien anglais de l'Imperial College, l'un des grands spécialistes

---

52. Albert Brester, *Essai d'une théorie du Soleil […]*, éd. cit., p. 7.
53. *Ibid.*, p. 19.

de spectroscopie, et Egon von Oppolzer (1869-1907), professeur d'astronomie à Innsbrück, l'un des fils du grand astronome autrichien Theodor von Oppolzer (1841-1886).

Suit un trou d'une dizaine d'années, jusqu'en 1907. Durant ces années, Brester prend connaissance de la découverte de l'électron et du phénomène de radioactivité et tente une nouvelle synthèse de sa théorie avec ces nouveaux éléments expérimentaux[54].

La troisième période de littérature concernant les travaux de Brester couvre les années de 1908 à 1925. Cette dernière époque est marquée par la littérature de langue française professionnelle : deux revues françaises – le *Bulletin astronomique* de l'Observatoire de Paris et la *Revue générale des sciences pures et appliquées* pour laquelle Brester est un collaborateur –, et la revue belge, *Ciel et Terre*, d'abord publiée par l'Observatoire royal de Belgique à Bruxelles à partir de 1880, puis reprise en 1910 par la Société royale belge d'astronomie[55].

C'est Alfred Fowler qui formule les critiques les plus précises et les plus importantes sur les travaux de Brester dès 1889 dans la revue *Nature*. Fowler attaque notamment le caractère généraliste de la théorie de Brester. Il fait remarquer que, selon le processus décrit, toutes les étoiles froides c'est-à-dire les étoiles rouges, devraient être variables ; or ce n'est évidemment pas le cas. Fowler, spectroscopiste reconnu à la fin du XIX[e] siècle, note que certaines étoiles rouges ne sont pas variables comme le montrent leurs spectres. Concernant les variations d'Algol, Fowler objecte que beaucoup d'étoiles montrent des caractéristiques spectrales identiques et qu'elles ne sont pas variables. Albert Brester, conscient que son approche n'explique pas tout, répond que dans le cas d'Algol (il refuse toute idée de compagnon tournant autour d'Algol), la variabilité est due à ses taches. En 1896, Fowler revient sur les observations spectroscopiques formulables à l'encontre de la théorie de Brester. En effet, les spectres de certaines étoiles rouges montrent des raies brillantes de l'hydrogène (raies en émission : figures 5 et 6). Or la théorie analogique et qualitative de Brester est incapable d'expliquer ce phénomène.

---

54. Voir la bibliographie de Brester en fin de chapitre : ouvrages de 1911 et de 1917.
55. Devenue depuis la Société royale belge d'astronomie, de météorologie et de géophysique du Globe.

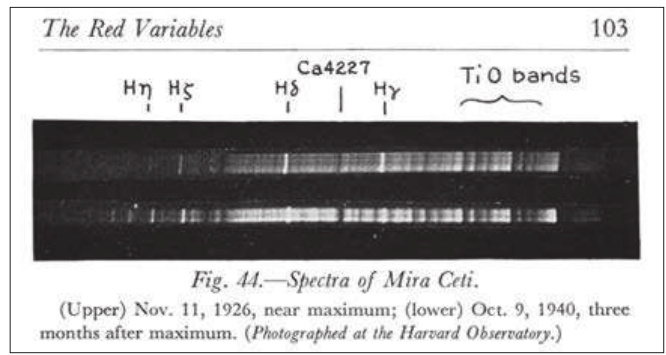

Figure 5. Exemples de spectres montrant les raies brillantes de l'hydrogène dans le spectre des étoiles rouges (de type M), par exemple la raie Hγ dans le spectre de l'étoile R Bootis à 4 340 ansgtröms. (Source : Paul W. Merril, *The nature of variable stars*, N. Y., McMillan Company, 1938, p. 78.)

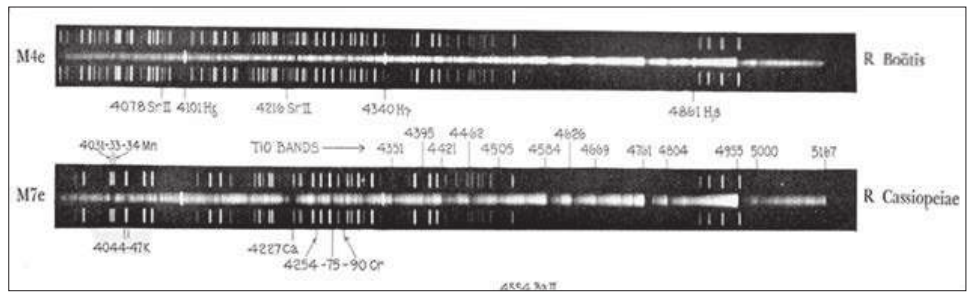

Figure 6. Spectre de Mira Ceti (1926) On voit ici comment lors des variations d'éclat, les raies d'émission de l'hydrogène dans le spectre de Mira Ceti, notamment les raies Hδ et Hγ, apparaissent ou disparaissent. (Source : L. Campbell et L. Jacchia, *The story of variable stars*, Philadelphia, The Blakiston Company, 1941, p. 103.)

Dans sa réponse, Brester fait alors appel au phénomène de luminescence [56] récemment proposé, qui selon lui, peut être produit par des probables chocs électriques à basse température dans l'atmosphère des étoiles rouges.

En 1895, von Oppolzer revient sur le calme de l'atmosphère solaire défendu par Brester. Oppolzer, farouche partisan de l'explication météorologique et des ouragans solaires lui objecte les forts mouvements de convection observés. Brester ne reconnaît pas ces mouvements violents et s'en tient à son hypothèse du Soleil tranquille. Ces prétendus mouvements violents sont pour lui des apparences de déplacements qu'il faut éclaircir à l'aide des nouvelles découvertes (mouvements électriques et thermiques dus aux déplacements d'électrons).

---

56. Eilhard Wiedemann, „Über Fluorescenz und Phosphorescenz, I. Abhandlung", *Annalen der Physik*, 34, 1888, p. 446-463. Le terme *Luminescenz* est introduit p. 447.

Regardons maintenant les relations que font les astronomes français et belges des travaux de Brester à partir de 1907. Si des critiques sont formulées dès la parution des premiers travaux de Brester, elles se font plus rares dans les années 1910 et la littérature comporte alors plutôt des éloges ou des appréciations très positives sur la théorie du Soleil et des étoiles variables que Brester tente de remettre en ordre et de promouvoir. Dans le *Bulletin astronomique*, revue de l'observatoire de Paris qui traduit les orientations scientifiques des astronomes parisiens, les avis portés sont assez positifs et l'on qualifie les idées de Brester, d'« ingénieuses » en attendant ses développements à d'autres domaines de la physique stellaire :

> […] Ces décharges intermittentes de l'énergie chimique que M. Brester appelle « éruptions de chaleur » seraient donc la cause des phénomènes périodiques stellaires. Ce sont elles qui, évaporant par intervalles les nuages obscurcissants, produisent les réhaussements d'éclat dans les étoiles variables. Ce sont elles qui, évaporant localement les nuages photosphériques du Soleil, y creusent des trous que nous apercevons comme taches. Ce sont elles aussi qui, en rendant lumineux les endroits où dans l'atmosphère tranquille du Soleil, les éléments jusqu'alors dissociés se combinent dès que leur perte continuelle de chaleur le permet, y allumant les protubérances et les rayons de la couronne […] M. Brester s'efforce de montrer que sa théorie explique les phénomènes solaires [et] se propose de la faire suivre d'une théorie des étoiles, des nébuleuses et des comètes, fondée sur les mêmes principes. Il sera alors plus facile de juger de la valeur de cette conception, d'ailleurs ingénieuse[57].

La revue belge *Ciel et Terre* présente souvent de manière élogieuse les travaux d'un des siens :

> M. Brester a depuis vingt ans, soutenu des idées nouvelles auxquelles on tend à se ranger aujourd'hui sur les causes des phénomènes offerts par la photosphère et ses enveloppes, couches renversante, chromosphère et couronne. Il a donné de très fortes raisons pour que l'on abandonne l'idée que les taches ou les protubérances sont dues à des mouvements de la matière gazeuse, soit circulaires soit de projection radiale, mouvements dont parfois la vitesse dépasse toute imagination. Il tend à n'y voir que la propagation de modifications physiques sans transport de matière […] L'impossibilité d'éruptions véritables dans l'atmosphère solaire […] est prouvée : 1°. Par la stratification durable des vapeurs solaires ; 2°. La rotation du Soleil par couches superposées qui tournent toutes autour de son axe avec des vitesses angulaires différentes ; 3°. Et surtout, l'invariabilité du spectre solaire. Voilà trois grands faits qui ont été mis en évidence par M. Brester dans ses nombreux travaux et dont les théories solaires doivent inéluctablement tenir compte […][58].

---

57. N.S., « Brester (A.) – Théorie du Soleil », *Bulletin astronomique*, série I, vol. IX, 1892, p. 426-427.

Lors de la parution de l'ouvrage posthume de Brester en 1924, Jean Mascart écrit une recension et explique que les éruptions de chaleur est un processus « qui n'est pas loin d'être accepté aujourd'hui ». Mascart signale « un recueil et un ouvrage du premier ordre » et souligne la richesse bibliographique de l'ouvrage qui recense près de 2 000 références aux théories solaires de 1840 à 1919.

Jean Bosler apprécie chez Brester des idées qui ont amené les astronomes à abandonner l'hypothèse météorologique. Après lui avoir consacré dix pages en 1910 dans son ouvrage *Les théories modernes du Soleil*[59], il en fait des éloges dans son cours d'astrophysique de 1928 :

Brester proteste avec raison contre n'importe quelle introduction de la météorologie terrestre dans la physique solaire ; L'immobilité du Soleil eut pu être seulement complète ; le spectrohéliographe montre des couches stratifiées et les lignes de Fraunhofer ne détectent aucun mouvement de la couche absorbante ; la dissociation doit être générale[60].

Bosler renforce sa bonne appréciation des idées de Brester et notamment l'idée des éruptions de chaleur :

Quand la température s'abaisse pour une raison ou pour une autre, les éléments dissociés se recombinent avec libération de leur chaleur de combinaison ; de même une condensation de vapeurs détermine un dégagement de chaleur de vaporisation. Les apparences de mouvements que nous croyons constater dans le soleil sont donc, trompeuses : la matière n'y bouge pas ; c'est son état chimique ou physique qui, seul, varie et sur place. Nous n'observons que des « éruptions » de chaleur. […] La photosphère n'est qu'une mer de nuages, séparés par des intervalles d'étendue très diverse qui sont les taches et les pores de la surface. Quant au rayonnement, il est inutile pour l'interpréter, d'avoir recours à des courants verticaux compliqués : la seule radiation des couches internes, se propageant par transparence à travers l'astre entier suffit. C'est d'ailleurs l'explication universellement adoptée aujourd'hui […][61].

---

58. E.L., « Le vulcanisme solaire de M. Krebs et la théorie solaire de M. Brester », *Ciel et Terre*, 33, 1912, p. 120-121, cit. p. 120.

59. Jean Bosler, « Théorie de Brester sur la dissociation », in *Les théories modernes du Soleil*, Paris, Gaston Doin, 1910, p. 62-71.

60. Jean Bosler, *Cours d'astronomie. III. Astrophysique*, Paris, Hermann, 1928, p. 204-205.

Notons une résurgence de la théorie de la dissociation dans les travaux du physicien Meghnad Saha (1921) soulignée par Bosler. En 1921, ce physicien montre que l'ionisation, c'est-à-dire, la décomposition d'un atome en un ion positif et des électrons est un phénomène de même nature que la dissociation réversible d'un gaz A en deux autres B et C (A = B + C)[62]. Edward Milne critiquera et complètera la même année cette idée dans un très intéressant article au style résolument moderne[63].

## CONCLUSION

Cette étude montre un très bon exemple de « second couteau » de la science, négligé à tort. Cette histoire met en relief les effets d'autorité de la « science des vainqueurs » (Lockyer, Young *et al.*) ou des « grands noms » de la science, qui occultent la manière dont la science se construit au jour le jour. La théorie solaire et généraliste de Brester du Soleil calme ou tranquille, constitue une étape intermédiaire entre les théories météorologistes et les théories modernes. Elle ne contient pas de mise en équation d'équilibres ou d'effets des rayonnements, mais elle comporte des idées réputées ingénieuses face auxquelles les concurrents ont à se positionner. La théorie solaire de Brester est une première tentative cohérente d'explication des variations des étoiles rouges variables qui oblige les théoriciens à préciser et compléter leurs arguments.

---

61. *Ibid.*
62. Jean Bosler, *op. cit.*, 1923, p. 33-34. Saha Meghnad, "Ionization in the Solar chromosphere", *Phil. Mag.*, vol. 40, october 1920, p. 472 ; "Problems of temperature radiation of gases", *Phil. Mag.*, vol. 41, february 1921, p. 267 ; "On a physical theory of stellar spectra", *Proc. Roy. Soc. A*, 99, 1921, p. 135. Voir aussi Anton Pannekoek, "Ionization in stellar atmospheres", *Bulletin of the Astronomical Institute of the Netherlands*, vol. 1, n° 19, 1922, p. 107-118.
63. Edward A. Milne, "Ionization in stellar atmospheres", *The Observatory*, vol. 44, n° 568, September 1921, p. 261-269.

L'étude de ces discussions montre et illustre les différentes attitudes des communautés scientifiques face aux développements de l'astrophysique et de la physique stellaire. Il est assez évident qu'à l'époque où les revues spécialisées en astrophysique voient le jour, les théories de Brester y trouvent leur place car les travaux les plus importants de physique stellaire ne sont pas encore connus ou sont encore en gestation. Pour Fowler et Von Oppolzer, la théorie de Brester constitue une curiosité qui ne les satisfait pas. Après 1896, les travaux de Brester ne seront plus discutés ni même mentionnés par les astronomes anglo-saxons, alors qu'ils constitueront une référence possible à ne pas écarter totalement chez certains astronomes français (ceux du *Bulletin astronomique* et les auteurs de la *RGSPA*) et ce, jusqu'en 1928 au moins (date de l'édition du cours d'astrophysique de Jean Bosler). Signalons toutefois que Jean Mascart et Jean Bosler sont les rares astronomes français à s'intéresser à la question des étoiles variables et à écrire sur ce sujet. Seul l'observatoire astronomique de Lyon a développé un programme d'observation et de réflexion théorique autour du problème des étoiles variables[64]. Il est donc assez normal de les voir discuter des théories d'Albert Brester.

Ce simple sondage bibliométrique et historiographique montre que le décalage théorique est patent entre les pratiques anglo-saxonnes et les pratiques franco-belges au début du XXe siècle. Dans les années 1920, il paraît évident que les astronomes français ne connaissent que peu ou pas les travaux d'Eddington, de Milne ou de Meghnad Saha et ne découvrent que peu à peu la nouvelle astrophysique qui s'est développée sans eux depuis les années 1890. N'est-il pas surprenant de lire chez Jean Bosler en 1928 que les idées de Brester sont sans doute une bonne voie ? Connaît-il à cette époque les nouvelles théories sur les intérieurs et pulsations stellaires d'Arthur Eddington[65] à l'esprit radicalement différent des théories « radio-calorifiques » de Brester ?

---

64. Voir la thèse de Michel Luizet, *op. cit.* ; articles parus dans le *Bulletin de l'Observatoire de Lyon*, 1920 : Henri Grouiller, « Le problème de la variabilité des céphéides », p. 81-83 ; *id.*, « L'observation visuelle des étoiles variables », (plusieurs parties), p. 145 *sq.* ; *id.*, « But des observations. Techniques particulières de l'observation des diverses catégories d'étoiles variables », p. 178 *sq.*, par exemple. Sur Michel Luizet (1866-1918) : « M. Luizet a eu un record honorable de 37 années de service à l'Observatoire, où il est entré à l'âge de 15 ans. De 1892 à 1911, il a eu la charge de la section météorologique, dans laquelle ses méthodes précises d'observation lui ont assuré le succès. En 1897, il a entrepris l'étude des Étoiles Variables, et ses nombreuses communications concernant ce sujet ont été publiées dans les *Comptes Rendus*, *Bulletin Astronomique*, […] ; il a également publié une importante monographie sur les étoiles variables Céphéides. Il a acquis aussi une place éminente parmi les observateurs de doubles étoiles, et était l'un des plus grands spécialistes français dans cette branche d'astronomie. » (*The Observatory*, janvier 1919.)

Cette histoire illustre aussi parfaitement pour l'historien la nécessaire prise en compte du temps nécessaire d'assimilation de nouveaux concepts scientifiques, de représentations et d'acceptation de phénomènes stellaires colossaux et démesurés en l'occurrence.

Signalons enfin qu'actuellement, on éprouve toujours quelques difficultés à interpréter les irrégularités dans les variations d'éclat et de période des étoiles rouges du type Mira Ceti, comme leur perte de masse (de $10^{-7}$ à $10^{-4}$ masse solaire par an). Plusieurs phénomènes semblent contribuer à ces irrégularités, à des parts inégales que l'on a bien du mal à mesurer, comme une savante combinaison d'ondes de chocs acoustiques propageant la chaleur vers les couches extérieures, de *thermal pulses* qui pourraient être assimilés aux « bouffées de chaleur » de Brester, d'abondances moléculaires inégales et variables révélées par la spectroscopie (présence de vapeur d'eau $H_2O$, monoxyde de carbone $CO$, silicates $SiO_2$, oxyde de titane $TiO_2$…) d'éjection de poussières dans les couches externes voire autour des étoiles, qui apparaissent depuis quelques années de surcroît dissymétriques dans les techniques d'imagerie stellaire par interférométrie infra-rouge[66]. Les atmosphères de ces étoiles rouges présentent toujours autant de mystères qu'à la fin du XIX[e] siècle.

---

65. Sir Arthur Eddington, *The internal constitution of the stars*, 1926 (réimpression Dover books, 1959) ; *id.*, « Theory of the Outer Layers of a pulsating star », *MNRAS*, 87/7, 1927, p. 539-553.

66. Donald G. Luttermoser, "The dynamic atmospheres of Mira variables stars", *Bull. Am. Astron. Soc.*, vol. 36, 2004, p. 817 *sq.* ; Markus Wittkowski *et al.*, "Inhomogeneities in molecular layers of Mira atmospheres", *Astron. & Astroph.*, 532, L7, 2011, p. 1-5. Voir aussi Keiichi Ohnaka, Thomas Driebe, Gerd Weigelt et Markus Wittkowski, "Temporal variations of the outer atmosphere and the dust shell of the carbon-rich Mira variable V Ophiuchi probed with VLTI/MIDI", *Astron. & Astroph.*, 466, 2007, p. 1099-1110.

# BIBLIOGRAPHIE CHRONOLOGIQUE D'ALBERT BRESTER JZ.

*Explication des phénomènes solaires les plus importants*, La Haye, W.P. Van Stockum and son, 1917.

*A summary of my theory of the sun*, La Haye, W.P. Van Stockum and son, 1919, 62 p.

*Le Soleil. Ses phénomènes les plus importants, leur littérature et leur explication*, La Haye, W.P. Van Stockum & Fils : ouvrage posthume édité par D<sup>r</sup> T. Van Lohuizen (professeur de physique à La Haye et la fille d'A. Brester, Nelly Brester), 1924, 315 p.

**Table 1. Recensions, critiques et réponses publiées concernant la théorie solaire d'Albert Brester, entre 1888 et 1920.**

| Année | Auteur | Revue et contenu |
|---|---|---|
| 1888 | | *Bulletin astronomique*, série I, n° 5, p. 549-550 : recension de l'*Essai d'une explication chimique des principaux phénomènes stellaires*. |
| 1889 | | *Bulletin astronomique*, série I, n° 6s, 496-497 : recension de l'*Essai d'une théorie du Soleil et des étoiles variables*. |
| 1889 | Fowler | *Nature*, 21 mars, p. 492 : recension et critique de l'*Essai d'une théorie du Soleil*… |
| 1889 | Brester | *Nature*, 25 avril, 606a, b : réponse de Brester |
| 1889 | Fowler | *Nature*, 25 avril, 606b : réponse de Fowler |
| 1892 | | *Bulletin astronomique*, série I, n° 9, p. 426-427 : recension de *Théorie du Soleil*. |
| 1894 | Brester | *Astronomy & Astrophysics*, vol. 13, décembre 1893, mars 1894, p. 218 : « A short review of my theory of the Sun ». |
| 1894 | Von Oppolzer | « On Brester's views as to the tranquility of the solar atmosphere », *Astronomy & Astrophysics*, 1894, p. 849-856 : réponse d'Egon von Oppolzer, sur la tranquillité de l'atmosphère solaire selon Brester. |
| 1895 | Von Oppolzer | *Astrophysical Journal*, 1, p. 260-262 : sur la tranquillité de l'atmosphère solaire selon Brester (même article que celui paru dans *A&A* en 1894). |
| 1895 | | *Nature*, 14 novembre : notice selon laquelle les raies de l'hydrogène devraient être produites à basse température, en accord avec la théorie de Brester. |
| 1896 | Brester | *Nature*, 16 janvier, p. 248-249 : réponse de Brester à la critique précédente et sur la variabilité des étoiles rouges. |
| 1909 | Brester via Julius | *PV de la section des sciences de l'Académie royale des Pays-Bas*, 11, 1908, p. 592-599 : critiques de Brester à la théorie des vortex de G. Hale. |
| 1909 | Brester | *RGSPA*, 20, p. 495-501 : « L'hypothèse des éruptions solaires » (qui doit être abandonnée). |
| 1909 | Brester | *RGSPA*, 20, p. 690-691 : Brester apporte de nouveaux éléments à propos de l'impossibilité des éruptions solaires. |
| 1911 | E.L. | *Ciel et Terre*, 32, p. 435-437 : Brester développe une approche du rayonnement radio-calorifique hypothétique émis par le Soleil. |
| 1912 | E.L. | *Ciel et Terre*, 33, p. 120-121 : comparaison du *vulcanisme solaire* de M. Krebs et de la théorie solaire de Brester. |
| 1917 | | *Monthly Weather Review*, octobre, p. 485 : sur la théorie solaire de Brester, revue pour la couronne regardée comme une aurore permanente maintenue par les électrons émis par la photosphère. |
| 1925 | Mascart | *RGSPA*, t. XXXVI, p. 84-85 : parution de *Le Soleil. Ses phénomènes les plus importants*… « Les travaux de Brester sont universellement connus et appréciés (84b). » |